# Lattice Vibrational Modes and Raman Scattering Spectra of Strained Phosphorene

*Ruixiang Fei and Li Yang\**

Department of Physics, Washington University in St. Louis, St. Louis, MO 63130, USA

KEYWORDS: Few-layer Black Phosphorus, Phosphorene, Strain, Raman Spectra.

ABSTRACT: Strain is prominent in fabricated samples of two-dimensional semiconductors and it also serves as an exploitable tool for engineering their properties. However, quantifying strain and characterizing its spatially inhomogeneous distribution across a material are challenging tasks. Here, we report the lattice vibrational modes and corresponding Raman spectra of strained monolayer black phosphorus (phosphorene) by first-principles simulations. We show that frequencies of vibrational modes of phosphorene and their Raman scattering peaks exhibit substantial and distinct shifts according to the types and size of strain. Therefore, combined with high spatial-resolution Raman scattering measurements, our calculated results can quantify strain distributions in phosphorene. This information is essential for understanding structures of future large-scale fabrication and strain engineering of phosphorene.

Strain is a deciding factor for numerous properties of semiconductors and can be used to tailor materials for novel applications.[1-7] This process is called strain engineering. Strain engineering is particularly effective for two-dimensional (2D) structures because these materials can sustain much larger strains than their bulk crystal counterparts.[8-11] In practice, the strain in a material is extremely complicated. For example, unlike the assumed uniform strain distribution used in most theoretical studies, any stretching or bending of 2D structures usually induces an inhomogeneous strain distribution. Moreover, large-scale samples from either exfoliations or epitaxial growth are known for fostering spatially inhomogeneous strains that strongly impact their device performance.[12,13] Therefore, measuring strain and, especially, characterizing its spatial distributions are major practical challenges for understanding experiments and performing strain engineering in 2D structures.

Recently, a novel type of 2D semiconductor, few-layer black phosphorus (phosphorene), was successfully exfoliated and has earned tremendous amounts of attention.[14-25] Strain and its spatial inhomogeneities are particularly pronounced in phosphorene due to two reasons: 1. Its unique puckered honeycomb structure shown in Figs. 1 (a) and (b) is easily distorted, resulting in strain. 2. Strain can dramatically modify a wide range of properties of phosphorene.[17, 19, 23] To extract strain information, lattice vibrational modes (phonons) and corresponding Raman scattering measurements are especially useful because they are directly decided by the atomic structure. In particular, current scanning confocal or near-field Raman scattering techniques can provide high spatial resolutions of local structures.[26-30] Ultimately, a first-principles prediction of lattice vibration modes and their Raman scattering spectra are the promising candidates for quantifying strain and its spatial distributions in phosphorene.

In this work, we employ the first-principles linear response method to calculate lattice vibrational modes and Raman spectra of strained phosphorene. We reveal the evolution of frequencies and Raman peaks of these phonon modes according to various strain conditions, including zigzag, armchair, and arbitrary biaxial ones. Substantial frequency shifts of these vibrational modes are observed and explained. Importantly, by examining Raman spectra, which can be conveniently measured, arbitrary strain information can be directly extracted from the strain-induced frequency shifts and variation in energy spacings between peaks of the characteristic vibrational modes. This approach paves the way for measuring strain and characterizing its inhomogeneous distributions, which are crucial for strain engineering phosphorene and general 2D semiconductors alike.

The top and side views of the atomic structure of phosphorene are presented in Figs. 1 (a) and (b), respectively. They are fully relaxed according the force and stress calculated using density functional theory (DFT) within the local density approximation (LDA) with an accuracy of 0.1 eV/Å. The Kohn-Sham equation is solved with normal-conserving pseudopotentials using a plane-wave cutoff energy of 35 Ry. The k-point sampling is set to be 14 x 10 x 1 for converged results.[18-20] The lattice vibrations and phonons are obtained by the linear response theory as implemented in Quantum Espresso.[31] Raman activities of vibrational modes at the zone center are derived from the widely used formulas[32]

$$I^{Raman} = 45(\frac{d\alpha}{dQ})^2 + 7(\frac{d\beta}{dQ})^2 = 45\alpha'^2 + 7\beta'^2 \qquad (1)$$

$$\alpha' = \frac{1}{3}(\tilde{\alpha}_{xx} + \tilde{\alpha}_{yy} + \tilde{\alpha}_{zz})$$
$$\beta'^2 = \frac{1}{2}[(\tilde{\alpha}_{xx} - \tilde{\alpha}_{yy})^2 + (\tilde{\alpha}_{xx} - \tilde{\alpha}_{zz})^2 + (\tilde{\alpha}_{yy} - \tilde{\alpha}_{zz})^2 + 6(\tilde{\alpha}_{xy} + \tilde{\alpha}_{yz} + \tilde{\alpha}_{zx})^2] \qquad (2)$$

where the Raman tensor is defined as $\tilde{\alpha}_{sl} = \left|\hat{e}_s \frac{d\tilde{\alpha}'}{dQ} \hat{e}_L\right|$, $\hat{e}_s$ and $\hat{e}_L$ are the polarization unit vectors of scattered and incident lights, respectively. These formulas consider the averaged result of all polarizations in scattering processes.

The phonon dispersion of monolayer phosphorene calculated by the linear response theory is presented in Fig. 1 (c). Correspondingly, we present atomic motions of vibrational modes in Fig. 1 (d). Two selection rules can be applied to decide their first-order Raman activities. 1) Only those optical modes at the Γ point (zone center) can be detected because of the small momentum of incident photons. 2) Only vibrational modes with the inversion symmetry can be Raman active based on the group theory.[33] Based on above criteria, $B_{1g}$, $B_{3g}^1$, $A_g^1$, $B_{3g}^1$, $B_{2g}$ and $A_g^2$ modes are Raman active while the other three optical modes, i.e., $B_{1u}$, $A_u$ and $B_{2u}$, are symmetry forbidden. We mark those Raman active modes with red and the inactive modes with black in Figs. 1 (c) and (d).

Next, we focus on the frequencies of lattice vibrational modes of strained phosphorene. There are two fundamental types of uniaxial strain, i.e., zigzag and armchair, in phosphorene. They are also the most interesting cases for strain engineering. On the other hand, biaxial strain can be regarded as the superposition of these two uniaxial strains and it will be discussed separately in the last part of this letter. As shown in Fig. 2, all of these vibrational modes exhibit substantial shifts under both types of uniaxial strain. Additionally, structural instabilities arise for armchair strain larger than - 4%. This can be seen from the negative frequencies of the $B_{1u}$ mode that occur in Fig. 2 (a). Therefore, we confine the range of armchair compression to be less than 4% in the following discussion.

In Fig. 2, frequency shifts are found to be correlated with atomic motions of modes, strain types and amplitudes. These shifts can be well explained by taking into consideration all of these factors. Let us take the $B_{2g}$ mode as an example. As shown in Fig. 1 (d), atomic motions associated with $B_{2g}$ occur mostly along the zigzag direction and its energy is chiefly decided by two motions, i.e., the anti-phase motion between atom 1 and atom 2 and the anti-phase motion between atom 2 and atom 3. Among them, the first one is the major factor because it contains more projected components of high-energy stretching and compression of the bond between atom1 and atom 2.

Given the above considerations, we can explain frequency shifts of the $B_{2g}$ mode. For the +2% zigzag stretch, our simulation shows that the bond length between atom 1 and atom 2 is stretched by around +2% as well. This enlarged bond length consequently reduces interatomic interactions, lowering the energy of the anti-phase motion between atom 1 and atom 2. Since this motion majorly determines the frequency of the $B_{2g}$ mode, we observe a significant red shift under zigzag stretch, as shown in Fig. 2 (c).

For the $B_{2g}$ mode under armchair strain, our first-principles simulation reveals that a +2% stretch mainly enlarges the angle between the bond connecting atoms 1 and 2 with the bond connecting atoms 2 and 3. Additionally, the stretching reduces the bond length between atoms 1 and 2 by 0.5%. This subsequently enhances the interatomic interactions between atom 1 and atom 2, resulting in a blue shift of this $B_{2g}$ mode. Meanwhile, because the variation of the bond length is smaller than that of zigzag strain, the size of the shift is substantially smaller, as seen in Fig. 2 (c). The above explanation based on variations of bond lengths can be employed to understand frequency shifts of most modes shown in Fig. 2.

These frequency shifts of strained phosphorene can be directly detected by Raman scattering spectra. As we discussed before, only lattice vibration modes $B_{1g}$, $B_{3g}^1$, $A_g^1$, $B_{3g}^1$, $B_{2g}$ and $A_g^2$ modes are Raman active. After calculating their Raman activities by Eqs. 1 and (2), only three modes, $A_g^1$, $B_{2g}$ and $A_g^2$, exhibit prominent Raman scattering peaks. In Fig. 3 (a), these peaks are located at 368, 433 and 456 cm$^{-1}$ for unstrained phosphorene, which are close to published experimental results.[20] The differences in peak positions and intensities between theory and experiments may result from three factors: 1. the inherent inaccuracies of first-principles DFT simulations; 2. the measured experimental samples may possess strain induced by defects or substrates;[15] 3. the details behind the scattering polarization setups may differ between theory and experiments since we assume the spatially averaged results in Eqs. (1) and (2). In light of this, we propose to extract the strain condition from the frequency shifts and energy spacings between peaks instead of their absolute positions and intensities. For this reason, the noted deviations will not significantly affect our theoretical predictions.

First, we focus on Raman spectra of phosphorene under zigzag strain, as shown in Fig. 3 (a). Different modes exhibit qualitatively different responses to applied strain. Both $B_{2g}$ and $A_g^2$ modes exhibit monotonic shifts; they are red shifted when stretched and blue shifted when compressed. The magnitudes of the shifts are prominent; a shift around 100 cm$^{-1}$ is observed for the $B_{2g}$ mode within a straining range of ±6%. It should be noted that the $A_g^1$ peak always has a slightly red shift (< 7 cm$^{-1}$), despite the sign of zigzag strain. Additionally, the energy spacing between them can be applied to justify the zigzag strain in phosphorene. For instance, the energy spacing between the $A_g^1$ and $B_{2g}$ modes decreases when the material is stretched but increases under compression. Interestingly, zigzag strain also substantially modifies the intensity of the Raman scattering signals. For example, the $B_{3g}^2$ signal is active but very weak in unstrained phosphorene but it is dramatically enhanced by four orders of magnitude under a +6% zigzag stretch as shown in Fig. 3 (a). This provides another characteristic feature for identifying zigzag strain in phosphorene.

For armchair strain, corresponding Raman spectra are presented in Fig. 4 (b). The main features are still dictated by three Raman active modes, $A_g^1$, $B_{2g}$, and $A_g^2$. However, they exhibit different characteristics than those in the zigzag case. All of them shift monotonically under armchair strain. However, in contrast to the zigzag strain case, $B_{2g}$ and $A_g^2$ show opposite shifting trends;

they exhibit red shifts under compression and blue shifts under stretching. $A_g^1$ is sensitive to the sign of the armchair strain and it shows a blue shift under compression and a red shift under stretching. Therefore, the energy spacing between $A_g^1$ and $B_{2g}$ modes increases under stretching and shrinks under compression. Another important feature is that the magnitude of frequency shifts of these modes is much smaller than in the case of zigzag strain. This is because phosphorene is rippled along the armchair direction and it is softer along this direction.

Beyond the uniaxial strain discussed above, biaxial strain is more widely observed in realistic samples on account of either epitaxial growth or external straining.[34, 35] More importantly, the biaxial strain is generally not uniform and may consist of different strain magnitudes along different directions. For example, it may be stretched more along the armchair direction than the zigzag direction even under same tension force. This is particularly relevant for phosphorene because its structure is anisotropic and much softer along the armchair direction. In order to capture these complicated biaxial straining conditions, we analyze the Raman peak positions of characteristic modes under arbitrary strain conditions.

Figure 4 reports the Raman peak position of those three characteristic modes, i.e., $A_g^1$, $B_{2g}$, and $A_g^2$, under arbitrary biaxial strain conditions. It is particularly noteworthy that these three modes exhibit very different responses under arbitrary strain conditions. For instance, the $B_{2g}$ is sensitive to zigzag strain but not to armchair strain. The $A_g^1$ mode is sensitive to both strain components and it varies significantly along the uniform biaxial strain. The $A_g^2$ mode is more sensitive to the zigzag strain but it can respond to the armchair strain under positive zigzag strain, as shown in Fig. 4 (c).

It must be pointed out that data points on the x (zigzag) or y (armchair) axis are not equal to the previously calculated zigzag or armchair uniaxial strain. In those cases, in addition to having an applied uniaxial strain, the structure is also fully relaxed along all other directions. For example, for a -2% armchair strain, we observe a 0.5 % contraction along the zigzag direction due to this relaxation. Therefore, the corresponding point is above the x axis, as marked by a star in Fig. 4 (c).

Figure 4 serves as a convenient chart for extracting arbitrary strain conditions of phosphorene. For example, we observe a 7 cm$^{-1}$ red shift of the $A_g^1$ peak and a 10 cm$^{-1}$ red shift of the $B_{2g}$ peak in a Raman scattering spectrum. Then we can mark corresponding isolines in Figs. 4 (a) and (b), respectively. The crossing point of these two isolines, which is marked by a cross in Fig. 4, indicates that the strain along the armchair direction is 1.3% and is 1.1% along the zigzag direction. Therefore, these figures can map those measured high spatial-resolution Raman spectra to strain conditions, which are crucial for understanding device performance and further strain engineering.

In conclusion, we have presented the lattice vibrational modes and corresponding Raman scattering spectra of strained phosphorene. Useful frequency shifts of the Raman activities of characteristic modes are revealed to be strongly correlated with strain conditions. By calculating these shifts under uniaxial and arbitrary biaxial strains, we provide a powerful tool for mapping the strain information from Raman measurements. This is important for understanding spatially inhomogeneous strain distributions and advancing strain engineering of phosphorene and general 2D semiconductors.

AUTHOR INFORMATION

**Corresponding Author**

* lyang@physics.wustl.edu

ACKNOWLEDGMENT

We acknowledge the fruitful discussions with Ryan Soklaski. This work is supported by the National Science Foundation (NSF) Grant No. DMR-1207141. The computational resources have been provided by the Lonestar of Teragrid at the Texas Advanced Computing Center (TACC).

**Figures and captions:**

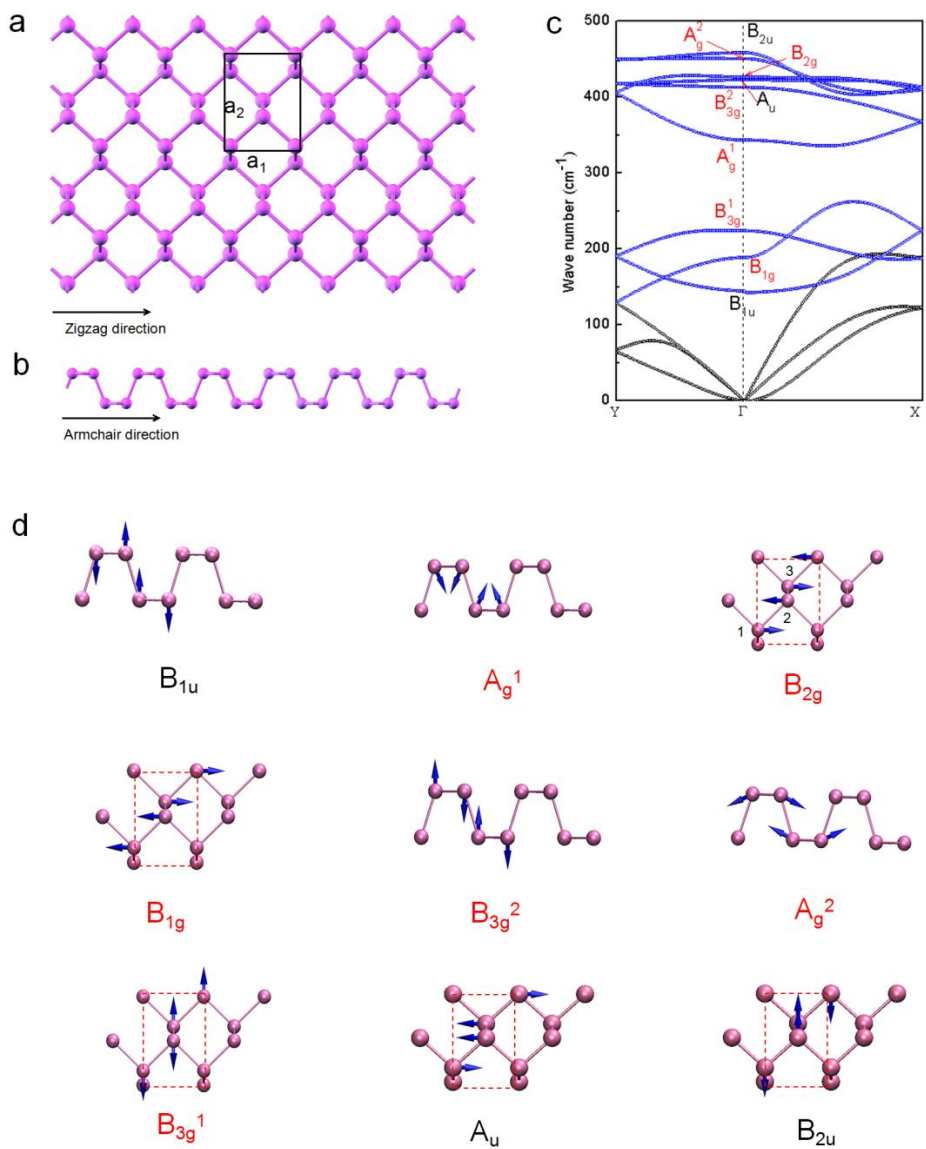

**Figure 1** (Color online) (a) Top view of the atomic structure of phosphorene. The unit cell and lattice vectors are marked. The zigzag and armchair directions are illustrated as well. (b) Side view of the phosphorene structure. (c) The phonon dispersion of unstrained phosphorene. (d) Atomic motions of lattice vibrational modes marked in (c). The view angle is specially selected to illustrate the most significant motions of vibrational modes. The length of the arrows represents the amplitude of motions. Raman active optical modes are marked by the red color while those inactive modes are marked by the black color.

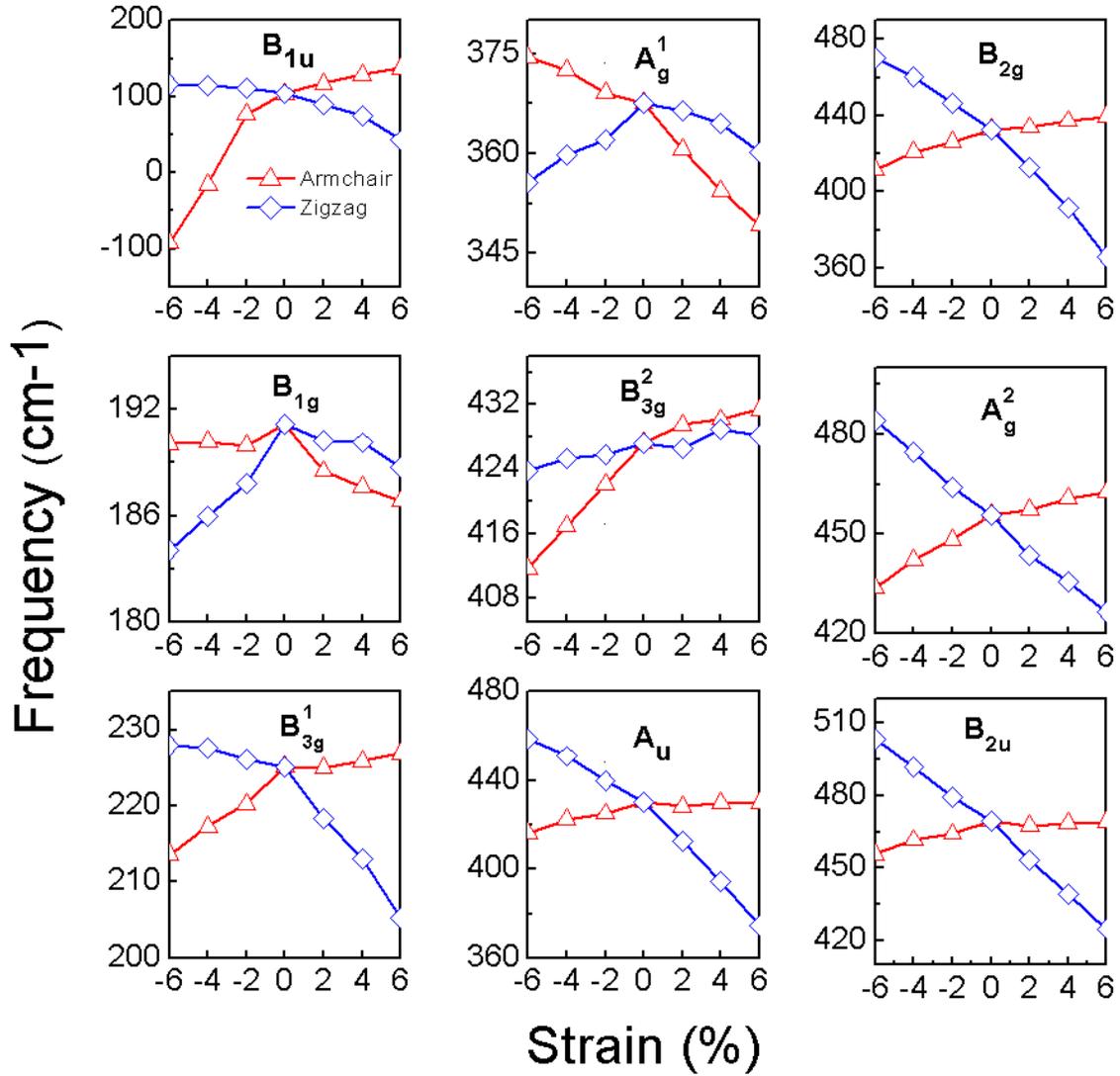

**Figure 2** (Color online) Frequencies of optical modes at the Γ point of phosphorene under uniaxial strain. Two typical types of uniaxial strain are included, i.e., zigzag and armchair. The negative frequencies observed in the $B_{1u}$ mode mean that the phosphorene structure is unstable under those strain conditions.

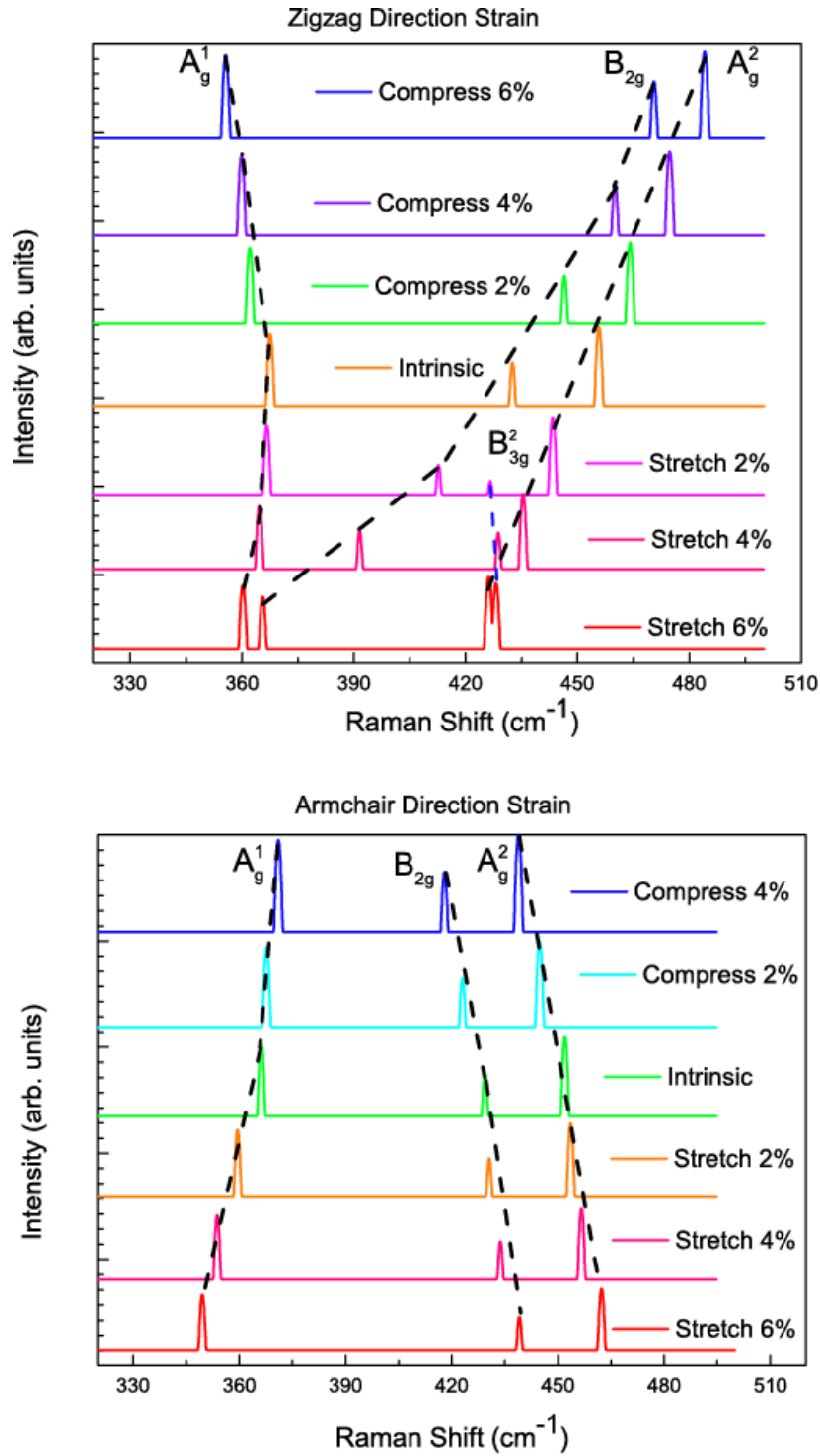

**Figure 3** (Color online) Raman spectra of phosphorene under uniaxial strain. (a) is about the zigzag strain and (b) is about the armchair strain. The peak height is in the logarithmic scale. We choose the arbitrary unit and the peak is with a 4 cm$^{-1}$ Gaussian smearing. The dashed lines are used to guide readers' eyes.

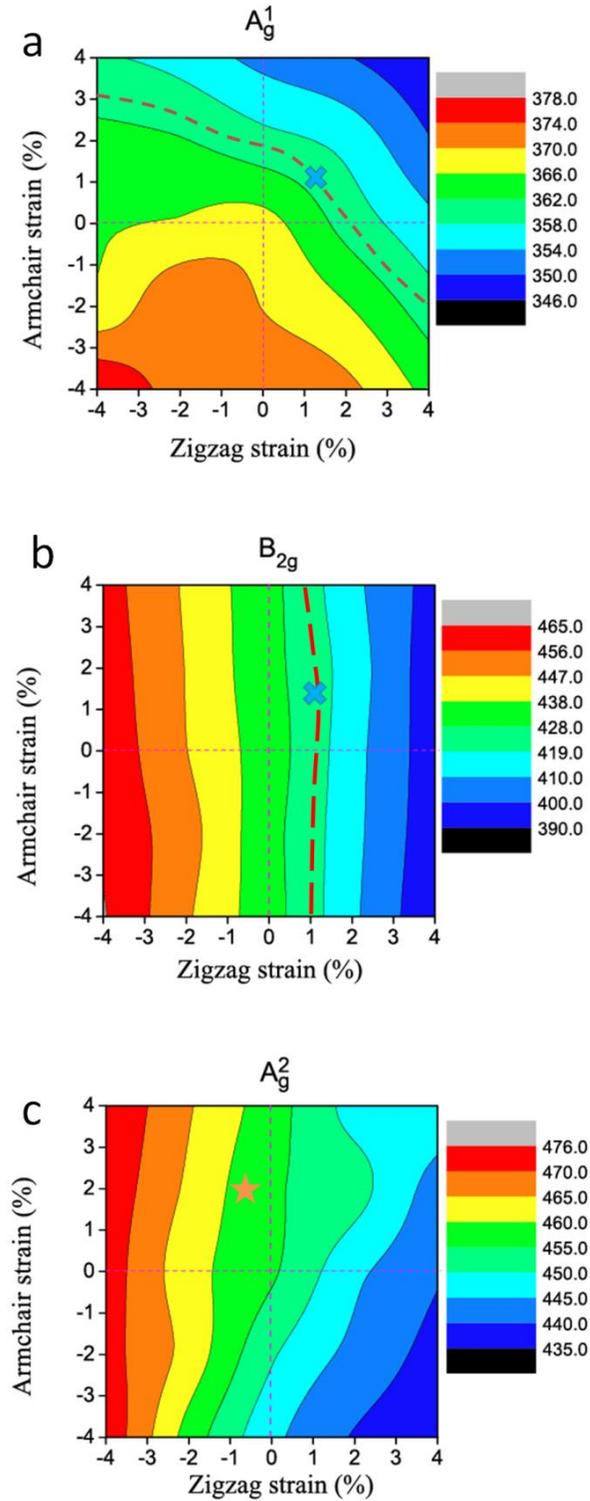

**Figure 4** (Color online) Color contour plots of Raman peak positions of the $A_g^1$ mode (a), the $B_{2g}$ mode (b), and the $A_g^2$ mode (c) of arbitrarily strained phosphorene. The blue colored cross in (a) and (b) represents the cross point of those dashed isolines in (a) and (b).